\documentstyle[prl,aps,epsf,floats]{revtex}
\begin{document}
\twocolumn[
\hsize\textwidth\columnwidth\hsize\csname@twocolumnfalse\endcsname
\draft
]
\noindent{\bf 
Comment on "Disorder and Quantum Fluctuations in Superconducting Films in Strong Magnetic Fields" :}

\bigskip

In a recent paper\cite{GaL}, Galitski and Larkin (GaL) have examined a 
macroscopic superconducting (SC) 
transition field $H_c$ at zero temperature ($T=0$) 
in disordered thin films under magnetic fields perpendicular to the plane. They have argued that $H_c$ usually lies {\it above} the ordinary mean field 
${\overline H}_{c2}(0)$ at $T=0$ and that an upwardly curved {\it nominal} 
$H_{c2}(T)$ curve defined from resistivity data\cite{Hebard,Mac} 
can be explained based on this $T=0$ result. 

Here, based on our recent works\cite{II,IAI}, we mainly focus on the situation denoted in Ref.1 as the mesoscopic disorder case and point out that the GaL's 
conclusion $H_c > {\overline H}_{c2}(0)$ contradicts available experimental 
data\cite{Okuma,Paa} suggesting a field-tuned superconductor-insulator 
transition (FSIT) and that this failure can be ascribed to their 
neglect of the amplitude-dominated quamtum superconducting (AQSC) 
fluctuation. When arguing that the transition dominated by the 
disorder-induced SC islands with local $H_{c2}$-values 
much higher than ${\overline H}_{c2}(0)$ can occur at $H_c$ much 
above ${\overline H}_{c2}(0)$, GaL have 
assumed that the AQSC fluctuation may be important only near 
${\overline H}_{c2}(0)$ so that an SC transition at $H_c$ may occur without 
being disturbed by the AQSC fluctuation. Since the correlator $C$ used in Ref.1 in defining $H_c$ diverges when the Edward-Anderson order parameter becomes 
nonvanishing, their $H_c$ should be identified with an FSIT field within the 
model in Ref.1. However, resistance data in 2D SC 
samples show a negative magnetoresistance\cite{Okuma,Paa} 
in higher fields than an (apparent) FSIT field, and this behavior is not seen \cite{Okuma} 
in 2D films nonsuperconducting even in zero field and in 3D SC 
samples. As pointed out elsewhere\cite{II}, this is best understood 
as the presence, {\it above} the (apparent) FSIT field, of 
nonvanishing fluctuation conductance terms\cite{GaL2} excluded from 
the Ginzburg-Landau description. Namely, 
the FSIT behavior occurs, contrary to the 
result in Ref.1, within or below the region around ${\overline H}_{c2}(0)$ in 
which the AQSC fluctuation is violent. 

In our opinion, an AQSC fluctuation peculiar to each of such islands with 
a higher $H_{c2}$-value begins to become important above 
${\overline H}_{c2}(0)$ consistently. A consistent treatment between the 
AQSC fluctuation and the vortex pinning effect has led to a $T=0$ FSIT field 
lower than ${\overline H}_{c2}(0)$ \cite{II,IAI}. 
We also note that, contrary to the data\cite{Hebard,Okuma,Paa}, GaL's 
eq.(16) results in an $H_c$ increasing with increasing disorder. 
According to Ref.5, an interplay between a microscopic 
disorder and an electron-electron 
repulsion needs to be incorporated to explain the FSIT field 
decreasing with increasing disorder. 

Regarding the resistive $H_{c2}(T)$ increasing upwardly upon 
cooling in 2D like systems\cite{Hebard}, 
we note that, if the SC fluctuation at measured temperatures is mainly not quantum but thermal in character, such an upward 
curve will be explained in terms of the SC fluctuation 
theory\cite{IOT} by phenomenologically incorporating a pinning strength through a random $T_c$. For 
instance, by defining a resistive $H_{c2}(T)$ in the manner 
$\rho(H=H_{c2}) = 0.9 \rho_n$, where $\rho$ and $\rho_n$ are, 
respectively, the total and normal resistivities, the resulting resistive 
$H_{c2}(T)$, in contrast to the GaL's strong disorder case, may 
lie below ${\overline H}_{c2}(T)$ and deviate 
upwardly at low enough $T$ from a nearly linear behavior \cite{Hebard} 
as a consequence of a decrease upon cooling of the SC fluctuation 
strength relative to the pinning strength. Further, if a resistive 
$H_{c2}(T)$ in 3D systems \cite{Mac} with strong SC 
fluctuation is defined as the positions at which the 
resistance (apparently) vanishes, it should also show an upwardly curved 
line below ${\overline H}_{c2}(T)$ reflecting the 3D vortex glass 
fluctuation created by the thermal or quantum SC fluctuation. 
Heat capacity data \cite{Mac2} and recent 
resistivity data \cite{Shiba} for overdoped cuprates strongly suggest 
${\overline H}_{c2}(T)$ lying far above the upwardly-curved\cite{Mac} resistive $H_{c2}(T)$. Similar remarkable differences between two nominal $H_{c2}(T)$ 
defined, respectively, from resistivity and other quantities are also found in electron-doped \cite{Nd} and (hole-)underdoped \cite{Capan} cuprates. Through such recent data in SC cuprates, it is believed that an 
upward resistive $H_{c2}(T)$-curve is not a reflection of the GaL's $T=0$ 
result in their strong disorder case but a direct consequence 
of SC fluctuation effects at nonzero $T$. 

\vspace{0.3truecm}
\bigskip
\noindent Ryusuke Ikeda \\
Department of Physics, \\
Kyoto University, Kyoto 606-8502, Japan


\vspace{-0.3truecm}
\begin{references}
\vspace{-0.3truecm}
\bibitem{GaL} V. M. Galitski and A. I. Larkin, Phys. Rev. Lett. {\bf 87}, 
087001 (2001). 
\bibitem{Hebard} See, for instance, A. F. Hebard and M. A. Paalanen, Phys. Rev. B {\bf 30}, 4063 (1984). 
\bibitem{Mac} A. P. Mackenzie et al., Phys. Rev. Lett. {\bf 71}, 1238 (1993). 
\bibitem{II} H. Ishida and R. Ikeda, J. Phys. Soc. Jpn. {\bf 71}, 254 (2002). 
\bibitem{IAI} H. Ishida, H. Adachi, and R. Ikeda, J. Phys. Soc. Jpn. {\bf 71}, 
245 (2002). 
\bibitem{Okuma} S. Okuma, S. Shinozaki, and M. Morita, Phys. Rev. B {\bf 63}, 054523 (2001). 
\bibitem{Paa} M. A. Paalanen, A. F. Hebard, and R. R. Ruel, Phys. Rev. Lett. {\bf 69}, 1604 (1992). 
\bibitem{GaL2} V. M. Galitski and A. I. Larkin, Phys. Rev. B {\bf 63}, 174506 (2001). 
\bibitem{IOT} R. Ikeda, T. Ohmi, and T. Tsuneto, J. Phys. Soc. Jpn. {\bf 60}, 1051 (1991). 
\bibitem{Mac2} A. Carrington, A. P. Mackenzie, and A. Tyler, Phys. Rev. B {\bf 54}, R3788 (1996).
\bibitem{Shiba} T. Shibauchi, L. Krusin-Elbaum, G. Blatter, and C. H. Mielke, unpublished. 
\bibitem{Nd} S. Kleefisch et al., Phys Rev. B {\bf 63}, R100507 (2001); F. Gollnik and M. Naito, Phys. Rev. B {\bf 58}, 11734 (1998). 
\bibitem{Capan} C. Capan et al., Phys. Rev. Lett. {\bf 88}, 056601 (2002). 

\end{references}
\end{document}